\documentclass{PoS}
\title{The density of states approach at finite chemical potential: a numerical study of the Bose gas.}

\ShortTitle{Short Title for header}

\author{\speaker{Roberto Pellegrini}\\
        Higgs Centre for Theoretical Physics, School of Physics and Astronomy, University of Edinburgh, Edinburgh EH9 3FD, United Kingdom\\
        E-mail: \email{r.pellegrini@ed.ac.uk}}

\author{Lorenzo Bongiovanni\\
  College of Science, Swansea University, Singleton Park, Swansea SA2 8PP, United Kingdom \\
        E-mail: \email{pylb@pyserver.swan.ac.uk}}

\author{Kurt Langfeld\\
  Centre for Mathematical Sciences, Plymouth University, Plymouth, PL4 8AA, United Kingdom\\
  E-mail: \email{kurt.langfeld@plymouth.ac.uk}}

\author{Biagio Lucini\\
  College of Science, Swansea University, Singleton Park, Swansea SA2 8PP, United Kingdom \\
        E-mail: \email{b.lucini@swansea.ac.uk}}

\author{Antonio Rago\\
  Centre for Mathematical Sciences, Plymouth University, Plymouth, PL4 8AA, United Kingdom\\
  E-mail: \email{antonio.rago@plymouth.ac.uk}}

\abstract{Recently, a novel algorithm for computing the density of states in statistical systems and quantum field theories has been proposed. The same method can be applied to theories at finite density affected by the notorious sign problem, reducing a high-dimensional oscillating integral to a more tractable one-dimensional one. As an example we applied the method to the relativistic Bose gas.}

\FullConference{The 33rd International Symposium on Lattice Field Theory\\
		14 -18 July 2015\\
		Kobe International Conference Center, Kobe, Japan*}

\begin{document}

\section{Introduction}
Non perturbative phenomena play a relevant role in many aspects of modern quantum field theory. The most developed tool to probe quantitatively these phenomena is the Markov Chain Monte-Carlo simulations of the theory discretised on a space-time lattice. However, this method has several limitations, one of the most severe being the so called sign problem.\\
The method relies on the interpretation of the Euclidean path integral measure as a probability measure; in which case it is possible to build a Markov Chain that will reproduce the Boltzmann weight in the long run. 
However, in many interesting theories the Euclidean action is not real and the method becomes very inefficient.\\
Recently, a new simulation method based on the density of states has been proposed \cite{LLR} and tested successfully in few models \cite{LLR_big,z3,heavyq,gattringer}.\\
We feel that the method deserves further testing in order to properly assess its performance.
So far, the method has been probed in discrete models and far from phase transitions. Our aim is to study a model with continuous degrees of freedom and in the neighboring of a phase transition. The relativistic Bose gas at finite density is very well suited for these kind of tests; it is known to undergo a second order phase transition and it is relatively fast to simulate.
\section{The density of states}
\begin{figure}[h]
    \centering
    \includegraphics[width=0.8\textwidth]{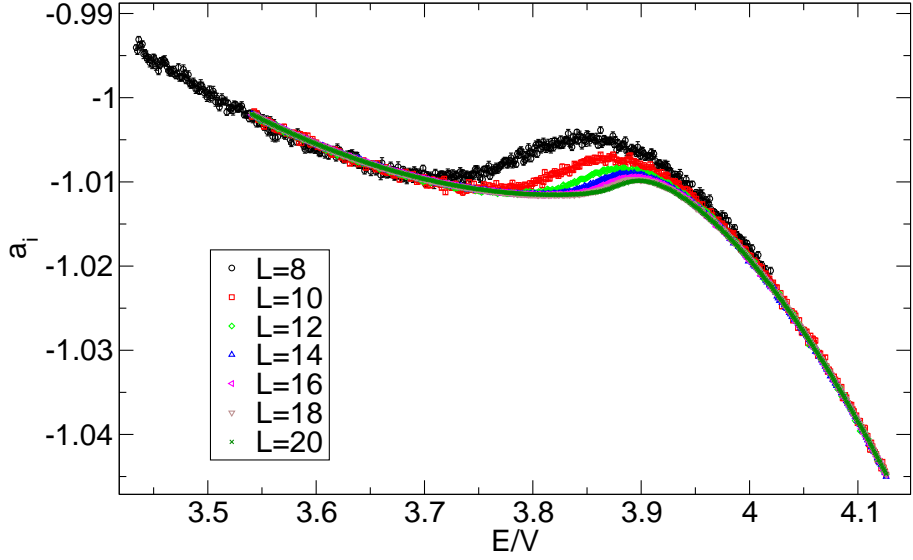}
    \caption{The log-derivative of the density of states for different volumes in 4d $U(1)$ LGT}
    \label{fig:rho_volume}
\end{figure}

Let us consider an Euclidean quantum field theory described by the variables $\phi$, the properties of the system can be derived from the partition function
\begin{equation}
  Z(\beta)=\int \left[ D\phi \right] e^{\beta S[\phi]}. \label{eq:partition}
\end{equation}
 The density of states is defined by the integral
\begin{equation}
  \rho(s)=\int \left[ D\phi \right] \delta(s-S[\phi]),
\end{equation}
and its geometrical interpretation is the volume of phase-space available to the system at fixed action.
From which it is possible to compute partition functions and observables
\begin{equation}
  Z(\beta)=\int ds \rho(s) e^{\beta s }.
\end{equation}
An algorithm to compute the log-derivative of the density of states was proposed in \cite{LLR}, and the convergence of the algorithm was proved in \cite{LLR_big}.\\
It was tested in 4d compact $ U(1) $ lattice gauge theory, which is known to undergo a first order phase transition with remarkable results.
In fig.(\ref{fig:rho_volume}) a plot of the log-derivative of the density of states is shown, and in tab.(\ref{tab:comparison}) we report
a comparison between the critical value of the coupling computed using the density of states method and a traditional Monte-Carlo.

\begin{table}
  \begin{center}
    \resizebox{5cm}{!}{
      \begin{tabular}{c c c c}
        $ L_{min} $ & $ k_{max} $ &  $ \beta_{C_v}(\infty) $ & $\chi^{2}_{red}$  \\
        \hline
        \textbf{14}  & \textbf{1} & \textbf{1.011125(3)} & \textbf{0.91}  \\
        \hline
        12 & 1 & 1.011121(3) & 2.42\\
        \textbf{12} & \textbf{2} & \textbf{1.011129(4)} & \textbf{0.67}\\
        \hline
        10 & 1 & 1.011116(5)& 7.44\\
        \textbf{10} & \textbf{2} & \textbf{1.011127(3)} & \textbf{0.60}\\
        \hline
        8 & 1 & 1.011093(5)& 90.26\\
        \textbf{8} & \textbf{2} & \textbf{1.011126(2)} & \textbf{0.62}\\
      \end{tabular}
    }
  \end{center}
\caption{Comparison between the critical value of $\beta $ computed using the density of states and using a traditional Monte-Carlo CITAZ}
\label{tab:comparison}
\end{table}
\section{Generalized density of states}
The case of complex action can be treated in the same fashion; eq.(\ref{eq:partition}) becomes
\begin{equation}
Z(\beta,\mu)=\int \left[D \phi \right] e^{\beta S_{R}[\phi]+ i \mu S_{I}[\phi]}
\end{equation}
where $ S_{R},S_{I}$ are the real and imaginary parts of the action and $\mu $ is the Lagrange multiplier correspondent to the complex part of the action. 
The generalized density of states \cite{z3} is introduced by
 \begin{equation}
P_{\beta}(s)=\int \left[D \phi \right] \delta \left(s-S_{Im}[\phi] \right) e^{\beta S_{Re}[\phi]}. \label{eq:generalized}
\end{equation}
If $ \beta=0 $, $P_{0}(s)$ is the volume of phase-space available to the system at given imaginary part of the action. At finite $\beta$ there is an additional weight proportional to $e^{\beta S_R[\phi]}$.
The partition function is the Fourier transform of the generalized density of states
\begin{equation}
  Z(\beta,\mu)=\int ds P_{\beta}(s) e^{i \mu s} \label{eq:complex_partition}.
\end{equation}
It is worth to notice that the integrand in eq.(\ref{eq:generalized}) is real and that the difficulties due to the oscillating phase are present only in the 1-dimensional integral in eq.(\ref{eq:complex_partition}).
Using the LLR algorithm it is possible to compute $P_{\beta}(s)$ with very high precision over many order of magnitude, thus the sign problem is reduced to the computation of a 1-dimensional fastly oscillating integral.
\section{Relativistic Bose gas at finite chemical potential}
The Bose gas at finite chemical potential is a perfect toy-model to test new algorithms for simulating systems affected by a sign problem.
It has been studied extensively with different approaches i.e. complex Langevin dynamics, Lefschetz thimble, dual formulation and mean field theory\cite{gattringerdual,Aarts1,Aarts2,forcrand}.\\
In the continuum formulation the action has the form
\begin{equation}
S[\phi]=\int d^4x\partial_{\mu} \phi \partial_{\mu} \phi+(m^2-\mu^2) |\phi|^2  + \mu (\phi^{*} \partial_4 \phi-\phi \partial_4 \phi^{*}) + \lambda  |\phi|^4 .
\end{equation} 
The discretization of this action poses no difficulties as long as the chemical potential is treated as a vector potential. This leads to
\begin{equation}
S[\phi]=\sum_{x} (2d+m^2) \phi^{*}_{x} \phi_{x} + \lambda (\phi^{*}_{x} \phi_{x})^2 -\sum_{\nu=1}^{4} \left( \phi^{*}_{x} e^{-\mu \delta_{\nu,4}} \phi_{x+\nu} + \phi^{*}_{x+\nu} e^{\mu \delta_{\nu,4}} \phi_{x}  \right).
\end{equation}
The latter can be written in term of the real and imaginary part of the fields $\phi_a=\frac{1}{\sqrt{2}} (\phi_1+i\phi_2 )$
\begin{equation}
S[\phi]=\sum_{x} (2d+m^2) \phi^{2}_{a,x} \phi_{x} + \lambda \phi^{4}_{a,x}  -\sum_{\nu=1}^{3} \phi^{a,x} \phi_{a,x+\nu} -\cosh{\mu}\phi_{a,x}\phi_{a,x+\hat{4}}+i \sinh{\mu} \epsilon_{ab} \phi_{a,x}\phi_{b,x+\hat{4}}.
\end{equation}
In the discrete case the Lagrange multiplier $\mu $ is coupled on both the real and imaginary part of the action, and as a consequence the generalized density of states will depend also on $ \mu $,
\begin{equation}
P_{m,\lambda,\mu}(s)=\int \left[D \phi \right] \delta \left(s-S_{Im}[\phi] \right) e^{S_{Re}(m,\lambda,\mu)[\phi]}
\end{equation}
We are particularly interested in the expectation value of the oscillating phase, which quantifies the severity of the sign problem
\begin{equation}
\langle e^{-S_{Im}} \rangle = e^{-V F}, \label{eq:f}
\end{equation}
where $ V $ is the volume and $ F $ is the variation of the free energy due to the oscillating phase.\\
Another interesting observable is the density of particles $\langle n \rangle $, given by
\begin{equation}
\langle n \rangle = \frac{d \log{Z}}{d \mu}.
\end{equation} 
\section{The oscillating integral}
\begin{figure}[h]
    \centering
    \includegraphics[width=0.8\textwidth]{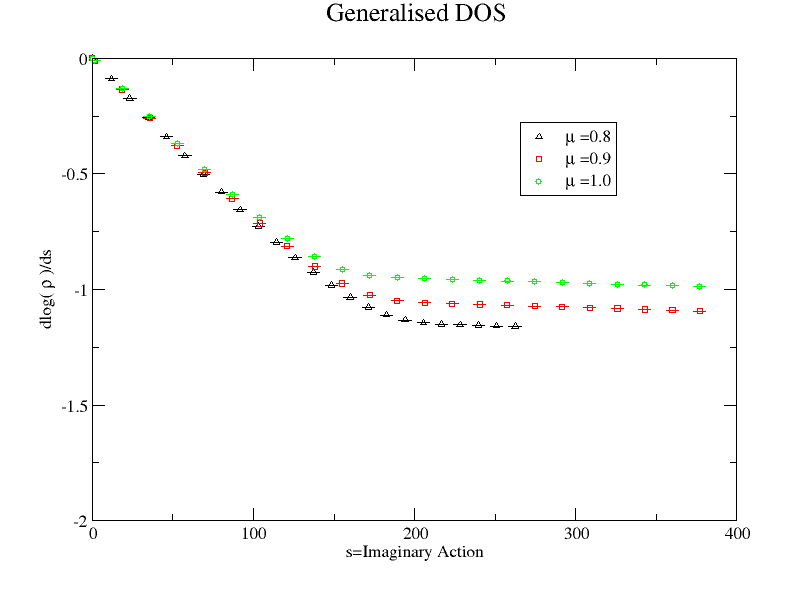}
    \caption{The log-derivative of the generalized DOS with $m=\lambda=1$ and $V=8^4$. A straight line would be a purely Gaussian DOS, it is very remarkable
that using the LLR is possible to clearly see deviations from gaussianity.}
    \label{fig:DOS}
\end{figure}

It is fairly easy to compute the generalized density of states using the LLR algorithm; an example is shown in fig.(\ref{fig:DOS}).
The LLR algorithm delivers a very high precision over different orders of magnitude. In fig.(\ref{fig:DOS}) it is clear that the DOS deviates from a pure Gaussian distribution. Deviations from a Gaussian function are clearly visible despite these are suppressed by a factor of order $e^{-100}$; which is well beyond the precision of a standard Monte-Carlo procedure. 

In order to extract physical quantities from the theory we need to carry out the Fourier transform of the generalized DOS, eq.(\ref{eq:complex_partition}).
The natural approach would be to use a fast Fourier transform algorithm, however this method is too naive and would break down at very low chemical potential.
In order to illustrate why this approach is not adequate let us assume that, in first approximation, the DOS is a Gaussian plus some white noise $\eta(s)$,
\begin{equation}
P_{m,\lambda,\mu}(s) \sim e^{-s^2 \sigma(m,\lambda)} + \eta(s).
\end{equation}
The partition function is the Fourier transform of the latter
\begin{equation} 
Z(m,\lambda,\mu) \sim  e^{-\frac{\mu^2 \pi^2}{ \sigma(m,\lambda,\mu)}} + c,
\end{equation}
where $ c $ is the Fourier transform of the noise. Therefore, the signal is exponentially suppressed with $\mu^2$ while the noise give a constant contribution to the partition function.
A much better alternative is to filter the noise by interpolating the log-derivative of the DOS with a polynomial of degree $2n$,
\begin{equation}
P_{m,\lambda,\mu}(s) \sim e^{\sum_{i} c_i s^{2i}}.
\end{equation}
In this case the statistical errors affect the coefficients and the Fourier transform of $P$ is a fast decaying function.
We found our results to be very stable whenever we avoid over-fitting; in practice we use a Bayesian evidence criterion \cite{Bayes} to decide the degree
of the polynomia in order to limit over-fitting.\\
Once a polynomial interpolant of the DOS is obtained, we are left with the computation of the following fastly oscillating integral
\begin{equation}
  Z(m,\lambda,\mu)=\int ds e^{\sum_{i} c_i s^{2i} +i \mu s}  \label{eq:complex_partition}.
\end{equation}
The most straightforward approach is to use a simple numerical integration routine with multiple precision.
The number of digits needed for the calculation increases with volume and chemical potential as the cancellations become more violent. It will eventually grow to a very large number. To give an idea, we needed 50 digits to compute the partition function with volume $8^4$ and $\mu=1$.\\
Another approach is the numerical steepest descent or Lefschetz Thimble \cite{Cristoforetti} method.
Let us consider an integral of the type
\begin{equation}
Z=\int_{-\infty}^{+\infty} e^{-S(x)} dx,
\end{equation}
where $S(x)$ is holomorphyic.
It is possible to show that
\begin{equation}
Z=\sum_{k} m_k e^{- i S_{Im}(z_k)} \int_{J_k} dz e^{-S_{Re}(z)},
\end{equation} 
here $z_k$ are critical points $ \partial_z S=0 $. And $\int_{J_k}$ are integrals over the curves of steepest descent.
These are parametric curves given by the ordinary differential equation
\begin{equation}
\dot{x}=-Re \left\{\partial_z S(z) \right\}, \ \ \dot{y}=+Im \left\{\partial_z S(z) \right\}.
\end{equation} 
$m_k$ is an integer number that counts the number of intersection between the curves of steepest ascent and the original domain of integration.
Notice that the sign problem did not disappear completely, there is a residual oscillating phase coming from the Jacobian of the transformation
$\frac{dz}{dt}$ and each curve of steepest descent contributes with a global phase $ e^{- i S_{Im}(z_k)} $. This residual sign problem is very mild and computation in double precision are sufficient. 
\begin{figure}
\begin{centering}
\includegraphics[width=.85\textwidth,height=.85\textwidth,keepaspectratio]{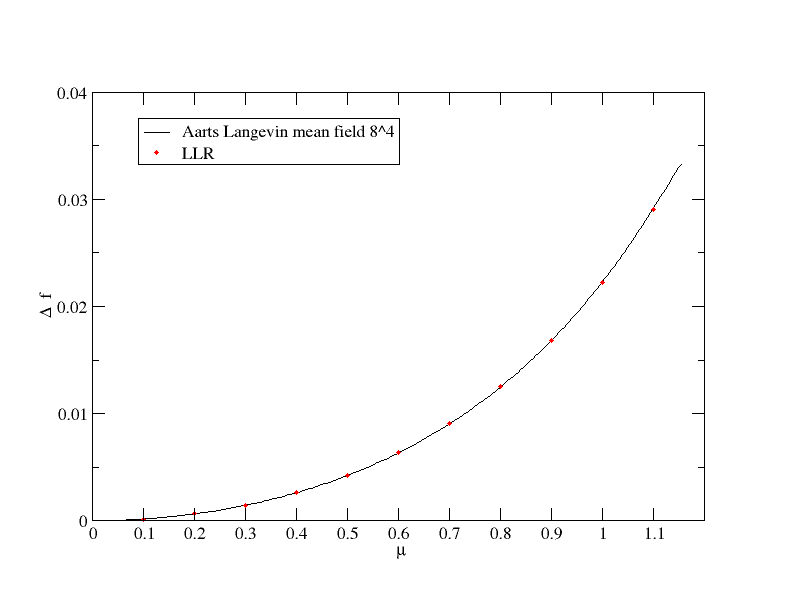}
\end{centering}
\label{fig:res}
\caption{Comparison between LLR method and mean field theory for $\mu=\lambda=1$}
\end{figure}
   
In Fig(\ref{fig:res}) we show the average free energy difference between the full and the quenched theory eq.(\ref{eq:f}). Our results are compared
with a mean field calculation in the framework of complex Langevin dynamics \cite{Aarts2}, courtesy of the author.\\
The results are in very good agreement with each other.
\section{Conclusions}
We presented an application of the LLR algorithm to a system affected by a severe sign problem.
We found that the method is able to extract meaningful results at finite density even in regions of the parameter space where the sign problem
is severe.  
The main drawback is that it relies on a polynomial fit of the log-derivative of the DOS, this might be difficult if the shape of the density of states is not very regular. On the other hand, in all the system studied with this approach the density of states turned out to be a remarkably smooth function, which gives hope that this is the case also for physically more relevant systems.

\end{document}